\documentclass[twocolumn]{article} 

\usepackage{graphicx}
\usepackage{hyperref}
\usepackage{twoopt}
\usepackage{authblk}
\newcommand{\ackname}{Acknowledgements}

\begin{document}

\title{Astronomy summer camp ``Beli Brezi'', Bulgaria -- building 
the astronomical community of the future}

\author[1]{Valentin D. Ivanov}
\author[2]{Agop Bohosian}
\affil[1]{European Southern Observatory, Karl-Schwarzschild-Str. 2,
85748 Garching bei M\"unchen, Germany; vivanov@eso.org}
\affil[2]{Astronomical Club ``Vega'', Kardzhali, Bulgaria;
obskar@gmail.com}
\date{}

\maketitle

\begin{abstract}
Why study astronomy, why teach astronomy? We give answers to 
these fundamental questions based on our experience with the 
Astronomical Camp ``Beli Brezi'' (White Aspens; Kardzhali, 
Bulgaria). It has been a place for teaching astronomy to high 
schools kids for nearly half a century. We describe shortly 
the history of the camp and draw some conclusions based on 
nearly five decades of experience. Major among them is that 
the camp has gone further than just distributing astronomical 
knowledge -- while this is an important and worthy task, the 
main achievement has been the cultivation of critical thinking 
among the pupils and we think that that is the main motivation 
to give positively reassuring answers the questions we asked 
at the beginning.
\end{abstract}
\vspace{1cm}

\section{Camp ``Bel Brezi'' -- an overview}

The longest-going astronomical summer camps in the Balkan 
peninsula ``Beli 
Brezi''\footnote{\url{http://www.astro-brezi.org/}} (White 
Aspens) started in the distant 1970. It is organized by the 
Astronomical Club 
``Vega''\footnote{\url{https://aovega.wordpress.com/}}, the 
Astronomy 
Department\footnote{\url{http://astro.phys.uni-sofia.bg/}} 
of the Physics Faculty at the University of Sofia and by the 
Astronomical Observatory ``Slavey Zlatev'', 
Kardzhali\footnote{\url{http://obskar.com/}}. The main goals of 
the camp are to train students in basic astronomy and various 
observing techniques, to give them an opportunity to collect 
observing data that will be processed and analyzed throughout 
the year, and to facilitate the interaction between the youth 
amateur-astronomer organizations across the country and in 
case of international participation -- from abroad.

The two-week long events usually take place in mid-Aug. Moon-phase 
permitting, they are 
timed to coincide with the peak of the Perseid meteor shower. The 
camp is opened to high school students with interest in astronomy, 
and with some background, typically from extra-curricular activities 
in their home high schools. Usually, the camp is attended by 40--60 
pupils. 

The instructors come from a number of backgrounds: a strong 
core is supplied by the Physics Department of the University of 
Sofia. Others are high school teachers and some work at 
the dozen or so public observatories scattered across the country. 
The last decade saw a remarkable new trend: former pupils, now 
Ph.D. students, postdocs or even professors/staff at various 
astronomical institutions abroad, return to teach.

The camps in the 1970-1990s were financed by the state's Department 
of Education. The main sponsor in the recent years has been the 
foundation {\it America for 
Bulgaria}\footnote{\url{https://www.us4bg.org/}}.

\section{Observing projects and equipment}

The type of observations carried out at the camp are typical for the 
amateur astronomers: the students were split into three teams: 
meteors, variables and photography.

Perhaps, the meteor observers are the most numerous. They measure
the hourly numbers and determine the radiants of the meteor showers
following closely the best practices adopted by the International 
Meteor Organization\footnote{\url{https://www.imo.net/}} (IMO).

Other students observe variable stars, both with naked eye and with 
binoculars or small telescopes (6-8\,cm refractors are the most
common). Observers estimate the apparent brightness of the stars with 
respect to reference stars following the classical methods of 
Argelander and Pickering and create light curves. Eclipsing binaries 
with sharp deep minima and some high-amplitude Cepheids are the most
common targets.

Finally, a third group carries out imaging of variable stars -- at
first by means of photographic film and plates, more recently with 
digital detectors. A decade ago these used to be long-term projects -- 
the students collected photographic material during the camp and 
process it on site, because
equipment like densitometers and blink-comparators were available 
only at the public observatories, for example at the one in Kardzhali.
These could only be used after the camp. The derived light curves 
and classifications of variables were typically reported at the 
annual national youth conferences that took place in early April the 
next year in the town of Varna. 

The extended duration of these projects forced the pupils to cultivate
persistence and to built planning skills. The digital era sped up this 
process and this has its advantages -- now the data analysis can be 
carried on site, and the pupils are introduced to and use 
professional-level software like IRAF\footnote{IRAF (Image Reduction 
and Analysis Facility) is distributed by the National Optical 
Astronomy 
Observatories, which are operated by the Association of Universities 
for Research in Astronomy, Inc., under cooperative agreement with 
the National Science Foundation.} to extract the photometry.
Naturally, the higher sensitivity of the electronic detectors 
allowed to image much deeper sky -- fainter galaxies, globular 
clusters. Attempts were made to expand the stellar variability work 
to transiting extrasolar planets. An amateur-level low-resolution 
optical spectrograph with a digital detector also became available 
to the pupils.

\begin{figure*} 
\centering
\includegraphics[width=16.6cm]{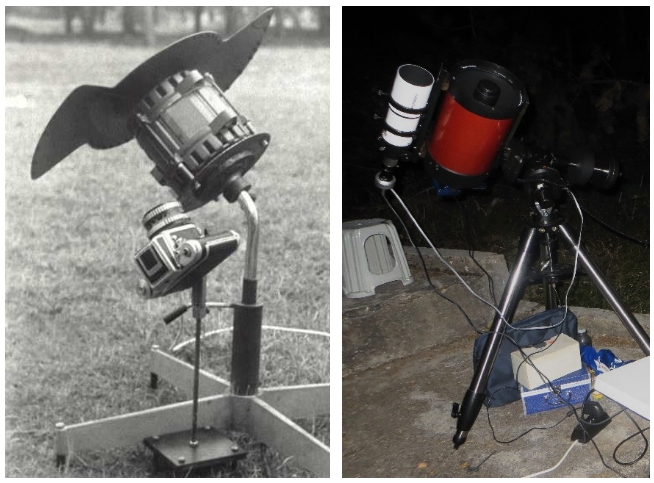}
\caption{Evolution of the equipment at the camp ``Beli Brezi'' over 
the last three decades: photographic camera with a student-built 
obturator for meteor observations from the mid-1980s (left) and a 
modern telescope with a digital camera used in 2016 (right). The 
number of telescopes -- most of them personal -- reached six at the 
2018 camp.}\label{fig:equipment}
\end{figure*}

Figure\,\ref{fig:equipment} shows example of the improvement in the 
equipment used at the camp over the last three decades.

\section{Discussion -- lessons learned from half-century of camp 
``Beli Brezi''}

Why teach astronomy? While we are at this, we may as well ask a 
more fundamental question -- who do astronomy at all? 

Indeed, astronomy was once an applied science: it produced calendars 
relevant for agriculture in the ancient times and in the age of the 
Great Geographic Discoveries finding your location and determining 
the time by astronomical means was a matter of survival at sea.
These times are long gone. Perhaps, the last astronomical discovery 
that had direct practical implications was that of helium in the 
solar spectrum. It happened in 1868, exactly a century and a half 
ago. 

The knowledge that we have gained about the Universe since then is 
enormous, but it has no direct applications in everyday life, unlike 
that yielded by other sciences like biology and chemistry. Of 
course, we can not exclude for sure that some day the astronomical 
research of today -- and may be even yesterday -- won't bring up 
some tangible improvements in humans' lives. Alas, that day has 
not arrived yet.

Surely, the astronomical research has had some indirect spin offs, 
e.g. development of advanced mathematical methods, optics and 
electronics, but in the research and development budgets for 
fundamental sciences like astronomy are far smaller than those 
for commercial and defence related companies and institutions. 
As a result astronomy is a recipient rather than a source of 
technological development.

Yet, the humanity keeps doing astronomical research -- albeit 
with a tiny fraction of the defence budgets: for example the 
annual budget of the largest astronomical organization in 
Europe -- the European Southern 
Observatory\footnote{\url{https://www.eso.org/}} (ESO) is 
comparable with the cost of a single modern fighter jet aircraft. 
There is only a handful of astronomical research institutions on 
the scale of ESO across the world; at the time of writing the 
combined fighter jet fleet all countries in the worlds numbers 
over 10,000 aircrafts of that 
type\footnote{\url{https://www.globalfirepower.com/aircraft-total-fighters.asp}}.
Assuming an average peace-time aircraft life of 20 years means that 
500 new machines are bought every year -- we give this example to 
add some hard numbers to our comparison.

It is not just the big astronomical organizations like ESO and 
the likes -- a handful of enthusiasts make sure the camp ``Beli 
Brezi'' -- and many other camps, summer or winter schools and other 
events -- keeps happening year after year. What motivates people to
volunteer their time and energy?

Let's look at the verifiable scientific output of the camp. As of 
2018 
ADS\footnote{\url{http://esoads.eso.org/abstract$\_$service.html}}
points at only a couple of Proceedings of the International Meteor 
Conferences \cite{Gavrilov_Chakarov_1995,Alexandrova_Stoilov_2010} 
that explicitly mention Beli Brezi in their abstracts. Furthermore, 
there is even a recent refereed publication by 
\cite{Kurtenkov_etal_2016}.
Perhaps, a more thorough search can identify some additional 
publications -- write ups for many more contributions from other 
meetings and conferences were never published as proceedings. These 
three (and hopefully other) publications are definitely a great 
achievement that can make the organizers of the camp proud. 

Another source of pride is the fact that many of the former pupils 
of the camp carry astronomical research now at places like the Center 
for Astrophysics at the Harvard University, the Flagstaff observatory, 
ESO, and at many leading universities -- Heidelberg, etc. These are 
exceptions at the level of a few percent from all pupils that have 
attended the camp, but great achievements nevertheless.

For the vast majority of the students, the encounter with the 
astronomical research early in their lives had stronger formative 
rather then educational impact. Self-discipline, organizational 
skills, sense of responsibility, teamwork, respect to others -- 
these are all qualities that the camp environment stimulates. 

Finally, and perhaps most importantly, the camp helped to propagate 
at the very basic level the habit of critical thinking among the 
pupils. For two weeks every year the students live in an environment 
where the scientific method is applicable by default and questions 
like: 

{\it Why do you think this way? 

How do you know? 

What is this statement based on?} 

are constantly asked back and forth. Our 
intention and hope is that the habit of asking such questions will
transfer into the everyday life of the pupils and that they would 
even ``infect'' with it other people around them and they will all 
apply it in everything -- from education at first to every time they 
have to vote or in their future jobs later on. Not surprising, many 
of the ex-students take up careers in STEM (science, technology, 
engineering and mathematics).

\section{Summary and conclusions}

The many years of camp ``Beli Brezi'' has though us that:
\begin{itemize}
\item the accessibility of science and the hands-on experience is 
critical for attracting young people;
\item camps like this help to propagate the critical scientific 
thinking throughout the society, this is our way to return the 
investment of public money into our science;
\item the modern incarnations of the camp rely heavily on the 
former alumni and many of them, who have become professional 
astronomers, are happy to come back as instructors;
\item international participation and connections help the students 
to experience the forefront of the modern astronomical science;
\item only a few alumni become professional astronomers, but a 
large fraction of them later build solid careers in STEM.
\end{itemize}

\section*{\ackname}
This is an extended write up of a poster presented at the European 
Week of Astronomy and Space Science (EWASS) held in Liverpool, Apr 
3-6, 2018, Special Session 8: Engaging the public with astronomy 
and space science research. We thank the organizers for giving us 
the opportunity to present the Astronomy camp ``Beli Brezi''.


\begin{thebibliography}{9}
\bibitem{Alexandrova_Stoilov_2010} 
Alexandrova, P.  \& Stoilov, B., Observations of the Perseids 2007 
meteor shower from Bulgaria during the National astronomy summer 
school in Belite Brezi. In Proceedings of the International Meteor 
Conference, 27th IMC, Sachticka, Slovakia, 2008, pages 10--15, 
August 2010.

\bibitem{Gavrilov_Chakarov_1995}
Gavrailov, A. \& Chakarov, R., Perseids 1994 - Observations from 
Beli Brezi. In A. Knoefel and P. Roggemans, editors, Proceedings 
of the International Meteor Conference, 13th IMC, Belogradchik, 
Bulgaria, 1994, 1995.

\bibitem{Kurtenkov_etal_2016}
Kurtenkov, A., Dimitrova, N., Atanasov, A. \& Aleksiev, T.~D. 2016,
RAA, 16, 105
\end{thebibliography}

\end{document}